\documentclass[nonacm]{acmart}

\usepackage{enumitem}       
\usepackage{booktabs}                       
\usepackage{array,ragged2e}                 
\newcolumntype{L}[1]{>{\RaggedRight}p{#1}}  
\newcolumntype{R}[1]{>{\RaggedLeft}p{#1}}   
\newcolumntype{C}[1]{>{\Centering}p{#1}}    
\newcolumntype{M}[1]{>{\RaggedRight}m{#1}}  
\newcolumntype{B}[1]{>{\RaggedRight}b{#1}}  
\usepackage{soul}           

\begin{document}

\acmBooktitle{arXiv}

\title{Adoption of smartphones among older adults and the role of perceived threat of cyberattacks}

\author{Patrik Pucer}
\email{patrik.pucer@fvz.upr.si}
\affiliation{
  \institution{University of Primorska}
  \city{Koper}
  \country{Slovenia}
}
\author{Boštjan Žvanut}
\email{bostjan.zvanut@fvz.upr.si}
\affiliation{
  \institution{University of Primorska}
  \city{Koper}
  \country{Slovenia}
}
\author{Simon Vrhovec}
\email{simon.vrhovec@um.si}
\affiliation{
  \institution{University of Maribor}
  \city{Maribor}
  \country{Slovenia}
}
\renewcommand{\shortauthors}{Pucer et al.}

\begin{abstract}
Adoption of smartphones by older adults (i.e., 65+ years old) is poorly understood, especially in relation to cybersecurity and cyberthreats. In this study, we focus on the perceived threat of cyberattacks as a potential barrier to smartphone adoption and use among older adults. The study also aims at investigating the differences between users and non-users of smartphones. We conducted a quantitative cross-sectional survey of older adults in Slovenia ($N = 535$). The results of covariance-based structural equation modeling indicate consistent support for the associations of intention to use (ItU) with perceived usefulness (PU), subjective norm (SN) and attitude toward use (AtU), the association between ease of use (EoU) and PU, the association between hedonic motivation (HM) and AtU, and the association between smartphone technology anxiety (STA) and fear of use (FoU). Even though the negative association between perceived threat (PT) and ItU was significant in the full sample, the non-user and the not aware subsamples, its role in adoption of smartphones among older adults remains puzzling. We uncovered significant positive associations between PT and AtU (except in the not aware subsample), and PT and PU which we could not fully explain in our study. The results of our study provide some insights on how campaigns promoting adoption of smartphones among older adults, workshops, training and informal teaching might be improved.
\end{abstract}


\maketitle

\section{Introduction}
\label{sec:intro}

Older adults (i.e., people aged 65 or older) represent a population that is growing rapidly worldwide. In EU countries, this number is increasing from 90.5 million at the start of 2019 to a projected 129.8 million in 2050 \cite{eurostat_ageing_2020}. With the global increase in the aging population, older adults are progressively using more and more different technologies, especially smartphones \cite{mihelic_use_2022}, driven by the desire to improve their quality of life and participate in various activities, such as entertainment, healthcare and social interaction \cite{huang_using_2023,yeh_effect_2024}. In the US, for example, over three quarters of older adults (76\%) own the smartphone \cite{bazen_cell_2023}. Hence, it is not surprising that non-use of smartphones in this population may lead to digital exclusion and unequal treatment \cite{zhu_staying_2024}.

This high rate of smartphone use by older adults poses a serious and subtle problem as it exposes this population to various cyberthreats which is particularly dangerous given the increasing number of older adults and the proportion of smartphone users in this population \cite{mccosker_60_2018,mihelic_secure_2022,tsatsou_aging_2021}. Older adults are also more vulnerable to cyberthreats due to declining cognitive abilities associated with aging \cite{mihelic_secure_2022}. Additionally, older adults face data and security issues with 49\% of mobile applications designed for them lacking any security or data protection measures \cite{portenhauser_mobile_2021}. Both users and non-users of smartphones and other smart devices among older adults often view inadequate security and data protection measures as a significant obstacle to their adoption and use \cite{harris_older_2022,kim_exploring_2021,kruse_mobile_2017,lamonica_understanding_2021,rasche_prevalence_2018}. They also lack awareness of information leakage and telecom fraud \cite{zhu_use_2023}. In the literature, there is a plethora of studies related to the adoption and use of smartphones by older adults. However, only a few studies investigate the issue of cybersecurity. There is a noticeable gap in the literature regarding the consideration of security concerns, particularly the perceived threat of cyberattacks, as a factor influencing the adoption and use of smartphones among older adults. This gap is significant because older adults’ perceptions of cyberthreats might significantly influence their willingness to adopt and use smartphones.

This study investigates the perceived threat of cyberattacks as a potential barrier to smartphone adoption and use among older adults. This study has three key objectives: 1) to study adoption of smartphones among older adults from new perspectives, 2) to study the role of perceived threat of cyber attacks in adoption of smartphones among older adults, and 3) to study the differences between users and non-users of smartphones. The study thus aims to provide a more comprehensive understanding of the factors that affect the adoption and use of smartphones in this demographic by considering not only the well-studied barriers, such as technology anxiety, but also the element of security concerns. The focus of this study is driven by the need to understand how concerns related to cybersecurity influence technology adoption among older adults. Our study also investigates the differences between older adults who use smartphones and those who do not, as well as the differences between those who are aware of smartphones and those who are not. These comparisons provide insights into the varying attitudes, behaviors, and challenges faced by older adults in the context of digital technology adoption.

\section{Theoretical background}
\label{sec:background}

\subsection{Technology adoption}

Our study strongly leans on the \textit{technology acceptance model} (TAM) \cite{davis_user_1989,venkatesh_model_1996} as the baseline research model. TAM aims to identify the key factors that influence new technology acceptance and helps to predict user behavior for different technologies and different user groups. Originally, TAM consisted of four key constructs: \textit{perceived usefulness} (PU), \textit{ease of use} (EoU), \textit{attitude toward use} (AtU), and \textit{intention to use} (ItU) \cite{davis_user_1989}. PU and EoU are determinants of AtU, and AtU and PU are determinants of ItU \cite{davis_user_1989}. TAM was later revised. Excluding AtU resulted in a more parsimonious TAM in which ItU is determined by PU and EoU \cite{davis_user_1989,lenz_why_2023}.

To overcome the limitations of the original and the revised TAM and to better understand the determinants of technology acceptance and use, Venkatesh and Davis \cite{venkatesh_theoretical_2000} developed TAM2 which provides a more sophisticated framework that takes into account the complexity of technology adoption in different contexts by providing a more comprehensive understanding of the factors that influence technology acceptance. Notably, \textit{subjective norm} (SN) was introduced into TAM2 even though it was purposefully excluded from TAM \cite{venkatesh_theoretical_2000}. SN is included in TAM2 as a predictor of both PU and ItU \cite{venkatesh_theoretical_2000}. Additionally, three more predictors of PU were introduced in TAM2: image, job relevance, and result demonstrability \cite{venkatesh_theoretical_2000}.

\subsection{Technology adoption by older adults}

Technology adoption by older adults is well studied. It should be noted that the literature defines varying age limits for older adults, the lower bound ranging from 50 to 65 years old. In this study we focus on older adults that are 65 or more years old. Nevertheless, we also reviewed the literature that defines older adults more broadly to cover a greater diversity of the published literature. Several theoretical models were used in the studies on technology acceptance, such as TAM \cite{conci_useful_2009,guner_use_2020,huang_what_2023,huang_using_2023,nayak_application_2010,ren_smartphone_2024,yeh_effect_2024,zhu_staying_2024}, TAM3 \cite{domingos_usability_2022}, the \textit{unified theory of acceptance and use of technology} (UTAUT) \cite{aranha_behavioural_2024,yang_acceptance_2023}, UTAUT2 \cite{huang_expanding_2023}, and the \textit{senior TAM} (STAM) \cite{chen_gerontechnology_2014,revollo_sarmiento_perceptions_2024}. These studies focused on adoption of different technologies and services, such as generic information and communication technology (ICT) \cite{guner_use_2020}, the internet \cite{nayak_application_2010} but also mobile phones \cite{conci_useful_2009,revollo_sarmiento_perceptions_2024}, smartphones \cite{huang_what_2023,huang_using_2023,ren_smartphone_2024,yeh_effect_2024,zhu_staying_2024,yang_acceptance_2023}, mobile applications and services \cite{aranha_behavioural_2024,huang_expanding_2023}, smart devices \cite{domingos_usability_2022}, and gerontechnology \cite{chen_gerontechnology_2014}.

The published literature identifies several relevant constructs as direct predictors of ItU technology by older adults. TAM constructs that have been found to be directly associated with ItU are PU \cite{conci_useful_2009,huang_using_2023,huang_what_2023,revollo_sarmiento_perceptions_2024,yeh_effect_2024} and its UTAUT alternative performance expectancy \cite{aranha_behavioural_2024,huang_expanding_2023,yang_acceptance_2023}, EoU \cite{conci_useful_2009,huang_using_2023,jarvis_technology_2020,revollo_sarmiento_perceptions_2024} and its UTAUT alternative effort expectancy \cite{aranha_behavioural_2024,huang_expanding_2023,yang_acceptance_2023}, and SN \cite{huang_using_2023} and its UTAUT alternative social influence \cite{conci_useful_2009,huang_expanding_2023,yang_acceptance_2023}. Although AtU was excluded from the revised TAM and its later versions, it was re-introduced in several studies among older adults \cite{chen_exploring_2021,guner_use_2020,huang_what_2023,jarvis_technology_2020,nayak_application_2010,ren_smartphone_2024,revollo_sarmiento_perceptions_2024,yeh_effect_2024,zhu_staying_2024}. Notably, some studies called for including both PU and AtU in studies among older adults \cite{yeh_effect_2024}. The results however appear inconsistent since some studies found none \cite{chen_gerontechnology_2014,revollo_sarmiento_perceptions_2024}, either PU \cite{revollo_sarmiento_perceptions_2024} or AtU \cite{guner_use_2020,nayak_application_2010,revollo_sarmiento_perceptions_2024}, or both being significant \cite{huang_what_2023,nayak_application_2010,yeh_effect_2024} which raises some questions about the mechanism of technology adoption in this population.

\subsection{The role of cybersecurity in technology adoption}

Adoption constructs can predict ItU smartphones in the older adult population relatively well \cite{cerda_diez_access_2023,yeh_effect_2024}. Nevertheless, some studies suggest that cybersecurity plays an important role in the adoption of ICT \cite{conci_useful_2009,lai_design_2016,mihelic_secure_2022}. This may be exacerbated by various issues, such as mobile apps for older adults with serious security flaws \cite{portenhauser_mobile_2021}.

Several constructs related to cybersecurity have been associated with adoption or actual use of ICT in the literature: perceived safety \cite{conci_useful_2009}, perceived security \cite{lai_design_2016}, perceived risk \cite{jena_factors_2023,zhao_what_2018}, and perceived trust \cite{jena_factors_2023,murko_bitcoin_2019}. Nevertheless, the \textit{protection motivation theory} (PMT) remains one of the key theories in the cybersecurity domain \cite{vrhovec_explaining_2023}. In the technology adoption domain, non-adoption of a technology can be considered as a protection measure which is conceptually in line with the technology threat avoidance theory (TTAT) \cite{liang_understanding_2010}. PMT includes two processes, namely threat and coping appraisal. According to PMT, individuals assess their vulnerability to a threat and its severity during threat appraisal which in turn influences their motivation to engage in protective actions \cite{burns_examining_2017,huang_expanding_2023,maddux_protection_1983}. There is support in literature for the mediating role of \textit{perceived threat} (PT) in associations of protection motivation with perceived vulnerability and perceived severity \cite{liang_understanding_2010,vrhovec_redefining_2021,fujs_social_2019}. Research suggests that PT may be a more accurate predictor of protection motivation than fear as individuals tend to rely on cognitive strategies rather than emotions when coping with threats \cite{johnston_speak_2019,vrhovec_redefining_2021}. In recent PMT studies, PT has been positively associated with both fear and protection motivation \cite{vrhovec_redefining_2021}, particularly in predicting older adults' fear of use of tablet computers \cite{mihelic_older_2023}. The literature shows mixed support for the role of cybersecurity in technology adoption. For example, perceived risk was found to be both significant \cite{jena_factors_2023,zhao_what_2018} and non-significant \cite{talukder_continued_2021,lutolli_adoption_2019} in the published literature. Similarly, perceived privacy has rarely been associated with adoption which can be attributed to the privacy paradox \cite{aranha_behavioural_2024,lenz_why_2023} -- the trade-off between losing a certain degree of privacy due to using a certain technology, and gaining some benefits from using it \cite{lenz_why_2023}. Nevertheless, information sensitivity seems to play a moderating role in technology adoption \cite{jelovcan_role_2020}. While perceived trust has been associated with adoption on technology \cite{talukder_continued_2021,murko_bitcoin_2019}, there is little support for a direct association between PT and ItU in the literature \cite{murko_bitcoin_2019}.

Only a few studies focused on older adults that were 65 or more years old which is the most widely accepted lower boundary for the age of older adults \cite{aranha_behavioural_2024,conci_useful_2009,frishammar_digital_2023,talukder_continued_2021,vrhovec_cybersecurity_2024}. A few more studies focused on older adults that were 60 or more years old (e.g., \cite{jena_factors_2023}) while the rest mostly focus on the general population (e.g., \cite{lai_design_2016,murko_bitcoin_2019}). These insights indicate that the relationship between technology adoption and cybersecurity constructs remains unclear, especially in the older adult population.

\section{Research model and hypotheses development}
\label{sec:model}

\begin{figure*}
  \centering
  \includegraphics[width=.75\linewidth]{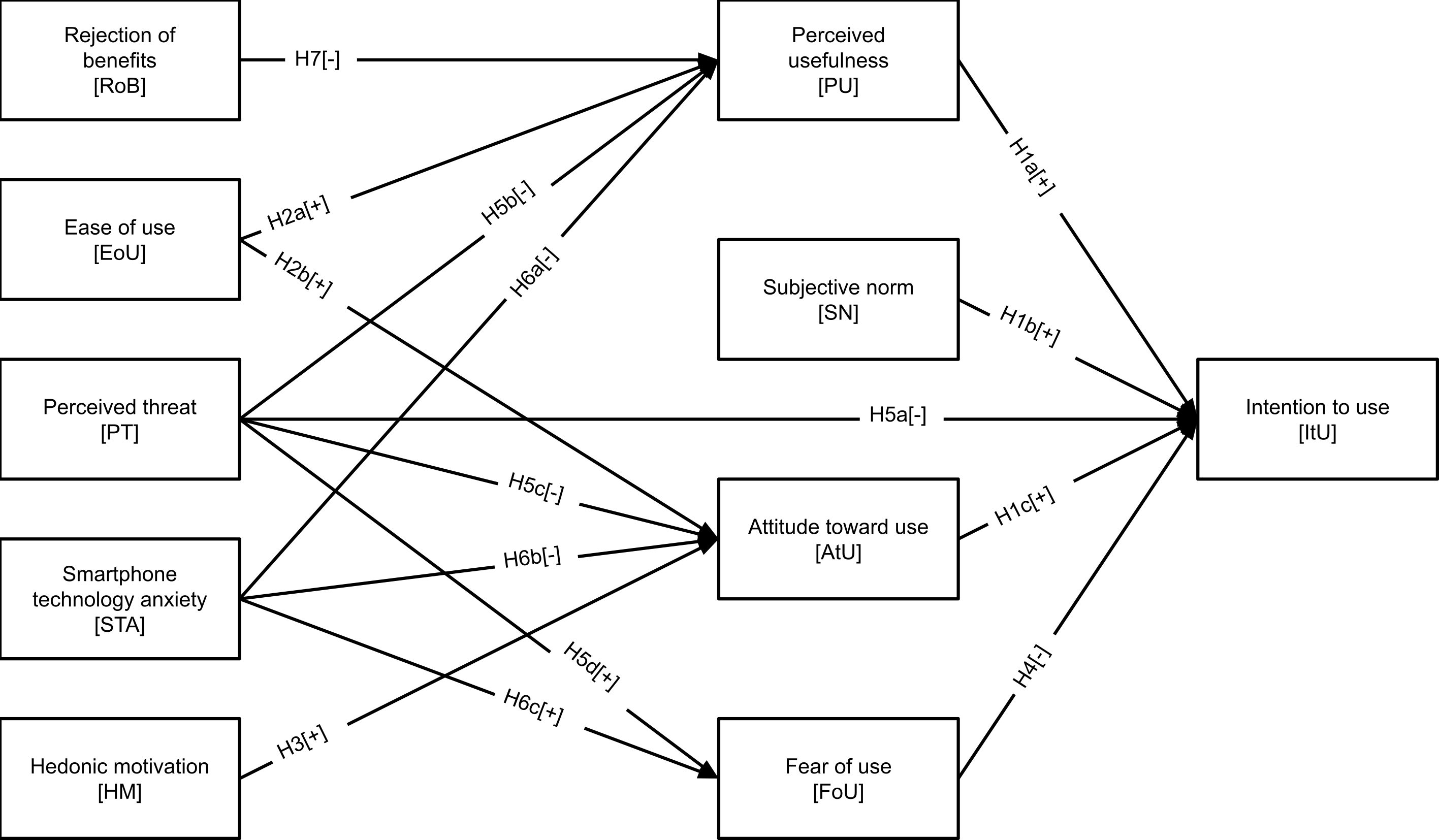}
  \caption{Research model.}
  \label{fig:rm}
\end{figure*}

The developed research model is presented in Fig.~\ref{fig:rm}.

According to TAM2, PU, SN and EoU are directly associated with ItU \cite{venkatesh_theoretical_2000}. The association between PU and ItU was confirmed among older adults in several studies, such as \cite{conci_useful_2009,huang_using_2023,huang_what_2023,mihelic_older_2023,yeh_effect_2024}. SN has also been found to have an association with ItU among the older adult population \cite{huang_using_2023}. Studies on older adults also indicate an association between AtU and ItU \cite{chen_exploring_2021,guner_use_2020,huang_what_2023,jarvis_technology_2020,yeh_effect_2024,zhu_staying_2024}. Even though AtU was excluded in favor of PU from the revised TAM and later versions \cite{davis_user_1989,lenz_why_2023}, there is some support for including both AtU and PU as predictors of ItU among older adults \cite{yeh_effect_2024}. Accordingly, we pose the following group of hypotheses:

\begin{itemize}[leftmargin=1.4cm]
    \item[\textit{H1a}:]{PU is positively associated with ItU.}
    \item[\textit{H1b}:]{SN is positively associated with ItU.}
    \item[\textit{H1c}:]{AtU is positively associated with ItU.}
\end{itemize}

The original TAM predicted associations of EoU with PU and AtU \cite{davis_user_1989}. While later versions of TAM, such as the revised TAM \cite{davis_user_1989} and TAM2 \cite{venkatesh_theoretical_2000}, exclude AtU from the model and directly associate EoU with ItU in addition to its association with PU, the results among older adults show poor support for these refinements. For example, there is mixed support for the direct association between EoU and ItU (significant \cite{conci_useful_2009,huang_using_2023,revollo_sarmiento_perceptions_2024}, non-significant \cite{chen_gerontechnology_2014,guner_use_2020,huang_what_2023,nayak_application_2010,revollo_sarmiento_perceptions_2024}). There is also support in the published literature for including AtU in research models when studying older adults \cite{yeh_effect_2024}, and associating it with EoU \cite{chen_gerontechnology_2014,guner_use_2020,huang_what_2023,jarvis_technology_2020,yeh_effect_2024}. There are however several studies supporting the association between EoU and PU as predicted by all versions of TAM \cite{chen_gerontechnology_2014,conci_useful_2009,guner_use_2020,huang_using_2023,huang_what_2023,jarvis_technology_2020,yeh_effect_2024,zhu_staying_2024}. Therefore, we develop the following set of hypotheses:

\begin{itemize}[leftmargin=1.4cm]
    \item[\textit{H2a}:]{EoU is positively associated with PU.}
    \item[\textit{H2b}:]{EoU is positively associated with AtU.}
\end{itemize}

UTAUT2 introduced \textit{hedonic motivation} (HM) as a predictor of ItU \cite{venkatesh_consumer_2012,lenz_why_2023}. HM is defined as the pleasure or fun stemming from using a technology, and is closely related to other constructs, such as enjoyment \cite{venkatesh_consumer_2012} and playfulness \cite{huang_what_2023}. There is poor support for a direct association between HM and ItU for older adults \cite{huang_expanding_2023}. Enjoyment has been associated with EoU and PU in studies on older adults \cite{conci_useful_2009}. Similarly, playfulness has been associated with PU but also AtU \cite{huang_what_2023}. We assume that HM primarily shapes someone's AtU hence we hypothesize:

\begin{itemize}[leftmargin=1.4cm]
    \item[\textit{H3}:]{HM is positively associated with AtU.}
\end{itemize}

Adoption of smart devices among older adults may be hindered by their fear that something might go wrong while using them \cite{mihelic_older_2023}. There are very few studies on the role of FoU in adoption of technology, especially among older adults \cite{mihelic_older_2023}. Therefore, we hypothesize:

\begin{itemize}[leftmargin=1.4cm]
    \item[\textit{H4}:]{FoU is negatively associated with ItU.}
\end{itemize}

PT has been directly associated with ItU \cite{liang_understanding_2010,fujs_social_2019,vrhovec_redefining_2021,burns_examining_2017} as well as fear \cite{vrhovec_redefining_2021,burns_examining_2017,vrhovec_explaining_2023}. Even though fear primarily entails the fear induced by a certain threat, such as cyberattacks, PT may induce a more specific kind of fear, namely FoU. Studies among older adults indicate support for this association \cite{mihelic_older_2023}. PT has also been indirectly associated with AtU through fear as a mediator \cite{vrhovec_explaining_2023}. These findings do not exclude a direct association between PT and AtU. Older adults may have a poorer AtU of smartphones if they perceive that they would be vulnerable to the threat of cyberattacks. PT may be similarly associated with PU since older adults may deem smartphones less useful if their use would make them prone to cyberattacks. Based on these considerations, we pose the following set of hypotheses:

\begin{itemize}[leftmargin=1.4cm]
    \item[\textit{H5a}:]{PT is negatively associated with ItU.}
    \item[\textit{H5b}:]{PT is negatively associated with PU.}
    \item[\textit{H5c}:]{PT is negatively associated with AtU.}
    \item[\textit{H5d}:]{PT is positively associated with FoU.}
\end{itemize}

STA has been associated with ItU in studies among older adults \cite{chen_gerontechnology_2014,huang_using_2023,huang_expanding_2023} even though the association is not always significant (e.g., \cite{chen_exploring_2021,ellis_why_2021}. We assume that the association between STA and ItU is indirect through AtU and PU. There is some support in existing literature for both associations with AtU \cite{chen_exploring_2021} and PU \cite{huang_using_2023,zhu_staying_2024}. Additionally, studies indicate that STA is associated with FoU \cite{mihelic_older_2023}. Thus, we develop the following set of hypotheses:

\begin{itemize}[leftmargin=1.4cm]
    \item[\textit{H6a}:]{STA is negatively associated with PU.}
    \item[\textit{H6b}:]{STA is negatively associated with AtU.}
    \item[\textit{H6c}:]{STA is positively associated with FoU.}
\end{itemize}

RoB appears to be relevant for a noticeable share of older adults however it has been rarely studied in the literature \cite{mihelic_older_2023}. By assuming that older adults who purposefully reject the benefits of smartphone use also perceive smartphones as less useful, we posit the final hypothesis:

\begin{itemize}[leftmargin=1.4cm]
    \item[\textit{H7}:]{RoB is negatively associated with PU.}
\end{itemize}

\section{Research methodology}
\label{sec:method}

\subsection{Research design}

We employed a cross-sectional research design to study the hypothesized associations. The survey was conducted among older adults (i.e., 65 or more years old) in Slovenia to determine factors associated with adoption to use smartphones (ItU), PU, AtU, and FoU. This research design also enabled us to determine differences between users and non-users of smartphones. The questionnaire was distributed in Slovenian which is the primary language of the studied population.

\subsection{Ethical considerations}

This study involved human participants. Therefore, the study proposal was submitted into consideration to an institutional review board. The study proposal was deemed as ethically acceptable and in compliance with the Code of Ethics and Integrity for Researchers at the University of Maribor by the Research Ethics Committee at the Faculty of Criminal Justice and Security at the University of Maribor on 27 February 2023 [2702A-2023]. Written informed consent was obtained from the participants.

\subsection{Measurement instrument}

Table~\ref{tab:constructs} presents the operational definitions of measured constructs. Most items were adapted to the context of our study from previously validated research. The three items for ItU were adapted from \cite{vrhovec_explaining_2023}, \cite{li_health_2019} and \cite{pan_internet_2010} each. Two items for PU were adapted from \cite{chen_gerontechnology_2014}, and the third one from \cite{zhu_staying_2024}. Two items for SN were adapted from \cite{macedo_predicting_2017}, one item from \cite{li_health_2019}, and one item was additionally developed for this study. One item for AtU was adapted from \cite{vrhovec_explaining_2023}, and two items from \cite{chen_exploring_2021}. Three items for EoU were adapted from \cite{macedo_predicting_2017}, and a fourth was additionally developed for this study. Items for HM were adapted from \cite{macedo_predicting_2017}. Two items for FoU were adapted from \cite{vrhovec_explaining_2023}, and a third one from \cite{moody_toward_2018}. Two items for PT were adapted from \cite{vrhovec_explaining_2023}, and a third one from \cite{liang_understanding_2010}. Items for STA were adapted from \cite{ellis_why_2021} while items for RoB were adapted from \cite{moody_toward_2018}. All items were measured on a 5-point Likert item scale from 1 \textit{strongly disagree} to 5 \textit{strongly agree}.

\begin{table*}
  \caption{Operational definitions of theoretical constructs}
  \label{tab:constructs}
  \centering
  \begin{tabular}{lL{.65\textwidth}}
    \toprule
Theoretical construct & Operational definition \\ \midrule
Intention to use [ItU] & The level of individual’s motivation to use or continue to use a smartphone. \\
Perceived usefulness [PU] & The degree to which an individual believes that using a smartphone will enhance their daily activities. \\
Subjective norm [SN] & The perception of social approval from important others regarding the use of a smartphone. \\
Attitude toward use [AtU] & An individual's positive or negative feelings about using smartphones. \\
Ease of use [EoU] & The degree to which an individual perceives that using a smartphone is free from effort. \\
Hedonic motivation [HM] & Fun or pleasure derived from using a smartphone. \\
Fear of use [FoU] & The level of an individual’s fear of using smartphones. \\
Perceived threat [PT] & The perceived extent of threats to the individual posed by cyberattacks. \\
Smartphone technology anxiety [STA] & The level of an individual’s anxiety related to the use of smartphones. \\
Rejection of benefits [RoB] & The degree to which an individual rejects the benefits of using a smartphone. \\
    \bottomrule
  \end{tabular}
\end{table*}

We measured the use of smartphones with the question \textit{Are you using a smartphone?} on a 3-point scale: \textit{Yes} (Use), \textit{No, but I have one} (Do not use), and \textit{No, and I don't have one} (Do not have). We measured awareness of smartphones with the question \textit{Before taking this survey, how aware were you of smartphones?} on a 4-point scale: \textit{I am very familiar with smartphones and what they are} (Very familiar with smartphones), \textit{I have heard about smartphones previously and am somewhat familiar with them} (Somewhat familiar with smartphones), \textit{I have heard mention of smartphones before but am largely unfamiliar with them} (Largely unfamiliar with smartphones), and \textit{I was not aware of smartphones before today} (Not aware of smartphones). This question was adapted from \cite{vrhovec_we_2024}.

\subsection{Data collection}

We collected data from 2 Mar 2023 to 19 Apr 2023. Respondents were recruited by students who were instructed to invite their grandparents, neighbors, acquaintances and random individuals from their hometowns to take part in the study. They were also instructed to help respondents with taking the survey if needed which involved making sure that the respondents understood the details about the survey that were provided on the survey introduction page, and the core concepts found in the survey (notably, smartphones and cyberattacks). The students were also trained in how to determine the most appropriate answer in case the respondents had trouble choosing an answer (e.g., agreement with a sentence on a 5-point scale). Each student recruited between one and 16 respondents. Students received a minor bonus in a course for their efforts.

We collected a total of 568 responses. After excluding respondents less than 65 years old, with missing or unrealistically high age (e.g., 194 years old), responses with over 10 percent missing values, and responses indicating respondent non-engagement, we were left with 535 useful responses which were used in further analysis. Sample characteristics are presented in Table~\ref{tab:sample}.

\begin{table*}
  \caption{Sample characteristics}
  \label{tab:sample}
  \centering
  \begin{tabular}{lrrr}
    \toprule
 & Frequency & Percent & Population \\ \midrule
\textit{Gender} &  &  &  \\
Female & 319 & 59.6\% & 56.4\% \\
Male & 216 & 40.4\% & 43.6\% \\
\textit{Age} &  &  &  \\
65-69 & 206 & 38.5\% & 30.5\% \\
70-74 & 129 & 24.1\% & 25.8\% \\
75-79 & 90 & 16.8\% & 17.2\% \\
80-84 & 62 & 11.6\% & 13.9\% \\
85+ & 48 & 9.0\% & 12.6\% \\
\textit{Formal education} &  &  &  \\
Finished high school or less & 349 & 65.2\% & 84.7\% \\
Achieved Bachelor’s degree & 124 & 23.2\% & 11.2\% \\
Achieved Master’s degree & 46 & 8.6\% & 3.5\% \\
Achieved PhD degree & 15 & 2.8\% & 0.7\% \\
N/A & 1 & 0.2\% &  \\
\textit{Living environment} &  &  &  \\
Urban & 259 & 48.4\% &  \\
Rural & 276 & 51.6\% &  \\
\textit{Smartphone use} &  &  &  \\
Use & 313 & 58.8\% &  \\
Do not use & 74 & 13.8\% &  \\
Do not have & 148 & 27.7\% &  \\
\textit{Awareness of smartphones} &  &  &  \\
Very familiar with smartphones & 174 & 32.5\% &  \\
Somewhat familiar with smartphones & 192 & 35.9\% &  \\
Largely unfamiliar with smartphones & 156 & 29.2\% &  \\
Not aware of smartphones & 13 & 2.4\% &  \\
    \bottomrule
  \end{tabular}
  \begin{flushleft}
  \textit{Note}: Demographic characteristics of the Slovenian older adult population in 2023 according to the Statistical Office of the Republic of Slovenia \cite{statistical_office_of_the_republic_of_slovenia_sistat_2024}.
  \end{flushleft}
\end{table*}

When comparing the demographic characteristics of our sample with the population \cite{statistical_office_of_the_republic_of_slovenia_sistat_2024}, we can see that it fits fairly well the population gender and age. Males are slightly overrepresented in our sample at the expense of females. Older adults 65-69 years of age are a tad overrepresented while older adults 85 years of age or older are slightly underrepresented. Older adults with the lowest levels of formal education are most underrepresented in our sample. This may be due to the recruiting strategy since students primarily recruited family members who may had a higher education than the population of older adults in Slovenia. When excluding older adults with finished high school or less, the shares of respondents who achieved Bachelor’s degree (67.0\%), achieved Master’s degree (24.9\%), and achieved PhD degree (8.1\%), are more comparable to the population of older adults in Slovenia (72.9\%, 22.6\%, and 4.5\%, respectively). Older adults who achieved Bachelor’s degree are slightly underrepresented, while older adults who achieved PhD degree are overrepresented. Overall, the respondents in our sample seem to have higher formal education than the population of older adults in Slovenia.

\subsection{Data analysis}

We used \textit{covariance-based structural equation modeling} (CB-SEM) to analyze the data. Data was analyzed with \textit{R} version 4.4.1 with packages \textit{lavaan} version 0.6-18 and \textit{semTools} version 0.5-6. Missing values (0.16 percent) were imputed with medians before analyzing the data with CB-SEM. We evaluated whether the data fit the measurement and structural models well with a set of model fit measures. We used \textit{comparative fit index} (CFI), \textit{standardized root mean square residual} (SRMR), \textit{root mean square error of approximation} (RMSEA). CFI values above the 0.95 threshold were considered excellent, and values above the 0.90 threshold were considered acceptable. SRMR values below the 0.05 threshold were considered excellent, and values below the 0.08 threshold were deemed acceptable. RMSEA values below the 0.06 threshold were deemed excellent, and values below the 0.10 threshold were considered acceptable.

Before testing the hypothesized associations, we conducted a \textit{confirmatory factor analysis} (CFA) to validate the measurement instrument. We assessed the reliability of the measurement instrument with \textit{composite reliability} (CR) and \textit{Cronbach’s alpha} (CA) coefficients. Values above the threshold value of 0.80 were considered excellent while values above the threshold value of 0.70 were considered acceptable. Convergent validity was assessed with \textit{average variance extracted} (AVE). Values above the threshold value of 0.50 were considered acceptable. Additionally, factor loadings above the 0.50 threshold provide further support for convergent validity. Discriminant validity was assessed with \textit{heterotrait-monotrait ratio of correlations} (HTMT) analysis. Values below the conservative 0.85 threshold are deemed acceptable. For conceptually similar theoretical constructs, values below the 0.90 threshold are considered acceptable.

After establishing adequate validity and reliability of the measurement instrument, we developed a structural model for testing the hypothesized associations. We calculated effect sizes $f^2$ for all hypothesized associations. To calculate effect sizes without changing the structural model, we calculated $R^2$ changes directly from the covariance matrix \cite{hayes_r-squared_2021}. Effect sizes above the 0.02, 0.15 and 0.35 thresholds were considered small, medium and large, respectively. The structural model was first applied to the full sample to test the hypothesized associations. Next, we split the sample into two subsamples, namely users and non-users of smartphones, and applied the structural model to both. To achieve this, we created two groups according to the use of smartphones. In the \textit{user} group, we included respondents who were using smartphones. In the \textit{non-user} group, we included respondents who either did not have a smartphone or had a smartphone but did not use it. This enabled us to compare differences in effect sizes and $R^2$ between users and non-users of smartphones. Finally, we split the sample according to the awareness of smartphones. We created two groups similar to \cite{vrhovec_we_2024} to improve the reliability of this single-item construct and the interpretability of the results. In the \textit{aware} group, we included respondents who were at least somewhat familiar with smartphones. In the \textit{not aware} group, we included respondents who were largely unfamiliar with smartphones or were not aware of smartphones before taking the survey. This enabled us to compare differences in effect sizes and $R^2$ between respondents who were aware of smartphones and those who were not.

\section{Results}
\label{sec:results}

\subsection{Instrument validation}

Model fit indices indicate that the measurement model fits the data well ($\chi^2 = 903.275$, $d. f. = 419$, $\chi^2 / d. f. = 2.156$, $CFI = 0.971$, $SRMR = 0.035$, $RMSEA = 0.046$). Measures for assessing validity and reliability of the measurement instrument are presented in Table~\ref{tab:validity}.

\begin{table*}
  \caption{Measurement instrument validity and reliability measures including HTMT analysis}
  \label{tab:validity}
  \centering
  \begin{tabular}{lrrrrrrrrrrrr}
    \toprule
Construct & CA & CR & AVE & 1 & 2 & 3 & 4 & 5 & 6 & 7 & 8 & 9 \\ \midrule
1: SN & 0.912 & 0.913 & 0.725 &  &  &  &  &  &  &  &  &  \\
2: PT & 0.888 & 0.895 & 0.743 & 0.119 &  &  &  &  &  &  &  &  \\
3: HM & 0.929 & 0.930 & 0.817 & 0.564 & 0.141 &  &  &  &  &  &  &  \\
4: EoU & 0.940 & 0.941 & 0.799 & 0.498 & 0.090 & 0.698 &  &  &  &  &  &  \\
5: STA & 0.872 & 0.872 & 0.695 & 0.383 & 0.106 & 0.504 & 0.533 &  &  &  &  &  \\
6: PU & 0.914 & 0.914 & 0.781 & 0.622 & 0.212 & 0.723 & 0.757 & 0.461 &  &  &  &  \\
7: ItU & 0.971 & 0.973 & 0.922 & 0.649 & 0.046 & 0.693 & 0.736 & 0.561 & 0.819 &  &  &  \\
8: FoU & 0.885 & 0.888 & 0.728 & 0.438 & 0.087 & 0.533 & 0.573 & 0.753 & 0.514 & 0.606 &  &  \\
9: RoB & 0.893 & 0.894 & 0.738 & 0.237 & 0.175 & 0.417 & 0.353 & 0.372 & 0.500 & 0.378 & 0.304 &  \\
10: AtU & 0.929 & 0.930 & 0.815 & 0.579 & 0.183 & 0.734 & 0.636 & 0.498 & 0.833 & 0.740 & 0.545 & 0.426 \\
    \bottomrule
  \end{tabular}
  \begin{flushleft}
  \textit{Notes}: CA -- Cronbach's alpha, CR -- composite reliability, AVE -- average variance extracted, HTMT -- heterotrait-monotrait ratio of correlations; SN -- subjective norm, PT -- perceived threat, HM -- hedonic motivation, EoU -- ease of use, STA -- smartphone technology anxiety, PU -- perceived usefulness, ItU -- intention to use, FoU -- fear of use, RoB -- rejection of benefits, AtU -- attitude toward use.
  \end{flushleft}
\end{table*}

CA and CR values ranged from 0.872 to 0.971 and from 0.872 to 0.973, respectively, suggesting excellent reliability. AVE ranged from 0.695 to 0.922 indicating adequate convergent validity. Also, item loadings ranged from 0.737 to 0.983 providing further support for this. Questionnaire items with loadings are presented in Table~\ref{tab:items}. HTMT analysis suggests that discriminant validity was adequate with values ranging from 0.046 to 0.833.

\begin{table*}
  \caption{Questionnaire items}
  \label{tab:items}
  \centering
  \begin{tabular}{lrL{.63\textwidth}l}
    \toprule
Construct & Loading & Item & Source \\ \midrule
SN & 0.843 & SN1. People who are important to me think that I should use a smartphone. & \cite{macedo_predicting_2017} \\
 & 0.850 & SN2. People whose opinions I value prefer that I use a smartphone. & \cite{macedo_predicting_2017} \\
 & 0.863 & SN3. My family and friends [support / would support] my decision to use a smartphone. & \cite{li_health_2019} \\
 & 0.853 & SN4. People who care for me [support / would support] my decision to use a smartphone. & Self-developed \\
PT & 0.737 & PT1. My smartphone [is / would be] threatened by cyberattacks. & \cite{vrhovec_explaining_2023} \\
 & 0.911 & PT2. Cyberattacks [are / would be] a danger to my smartphone. & \cite{vrhovec_explaining_2023} \\
 & 0.914 & PT3. Cyberattacks [pose / would pose] a threat to my smartphone. & \cite{liang_understanding_2010} \\
HM & 0.920 & HM1. Using a smartphone [is / would be] fun. & \cite{macedo_predicting_2017} \\
 & 0.889 & HM2. Using a smartphone [is / would be] enjoyable. & \cite{macedo_predicting_2017} \\
 & 0.902 & HM3. Using a smartphone [is / would be] entertaining. & \cite{macedo_predicting_2017} \\
EoU & 0.891 & EoU1. I find smartphones easy to use. & \cite{macedo_predicting_2017} \\
 & 0.895 & EoU2. My interaction with smartphones [is / would be] clear and understandable. & \cite{macedo_predicting_2017} \\
 & 0.857 & EoU3. Learning how to use a smartphone [is / would be] easy for me. & \cite{macedo_predicting_2017} \\
 & 0.927 & EoU4. I can easily handle smartphones. & Self-developed \\
STA & 0.759 & STA1. Using a smartphone makes me nervous. & \cite{ellis_why_2021} \\
 & 0.874 & STA2. Smartphones [make / would make] me feel uncomfortable. & \cite{ellis_why_2021} \\
 & 0.870 & STA3. I get a sinking feeling when I think of using a smartphone. & \cite{ellis_why_2021} \\
PU & 0.897 & PU1. Using a smartphone [makes / would make] my life more convenient. & \cite{chen_gerontechnology_2014} \\
 & 0.897 & PU2. I [find / would find] a smartphone useful in my life. & \cite{chen_gerontechnology_2014} \\
 & 0.857 & PU3. Using a smartphone [makes / would make] my daily activities easier to do. & \cite{zhu_staying_2024} \\
ItU & 0.983 & ItU1. I will use a smartphone in the near future. & \cite{vrhovec_explaining_2023} \\
 & 0.979 & ItU2. I intend to use a smartphone in the near future. & \cite{li_health_2019} \\
 & 0.915 & ItU3. I am interested in using a smartphone in the near future. & \cite{pan_internet_2010} \\
FoU & 0.893 & FoU1. I am afraid of using a smartphone. & \cite{vrhovec_explaining_2023} \\
 & 0.844 & FoU2. Using a smartphone is frightening.  & \cite{vrhovec_explaining_2023} \\
 & 0.810 & FoU3. Something terrible will happen if I use a smartphone. & \cite{moody_toward_2018} \\
RoB & 0.838 & RoB1. The benefits of using smartphones are not realistic. & \cite{moody_toward_2018} \\
 & 0.876 & RoB2. The benefits of using smartphones are overly exaggerated. & \cite{moody_toward_2018} \\
 & 0.860 & RoB3. The benefits of using smartphones are overstated. & \cite{moody_toward_2018} \\
AtU & 0.883 & AtU1. Using a smartphone is beneficial. & \cite{vrhovec_explaining_2023} \\
 & 0.913 & AtU2. Using a smartphone is a good idea. & \cite{chen_exploring_2021} \\
 & 0.911 & AtU3. I like the idea of using a smartphone. & \cite{chen_exploring_2021} \\
    \bottomrule
  \end{tabular}
  \begin{flushleft}
  \textit{Notes}: SN -- subjective norm, PT -- perceived threat, HM -- hedonic motivation, EoU -- ease of use, STA -- smartphone technology anxiety, PU -- perceived usefulness, ItU -- intention to use, FoU -- fear of use, RoB -- rejection of benefits, AtU -- attitude toward use.
  \end{flushleft}
\end{table*}

The results of CFA indicate that the measurement instrument adequately fits all criteria for establishing its validity and reliability which enables us to proceed to hypothesis testing.

\subsection{Hypothesis testing}

\begin{figure*}
  \centering
  \includegraphics[width=.99\linewidth]{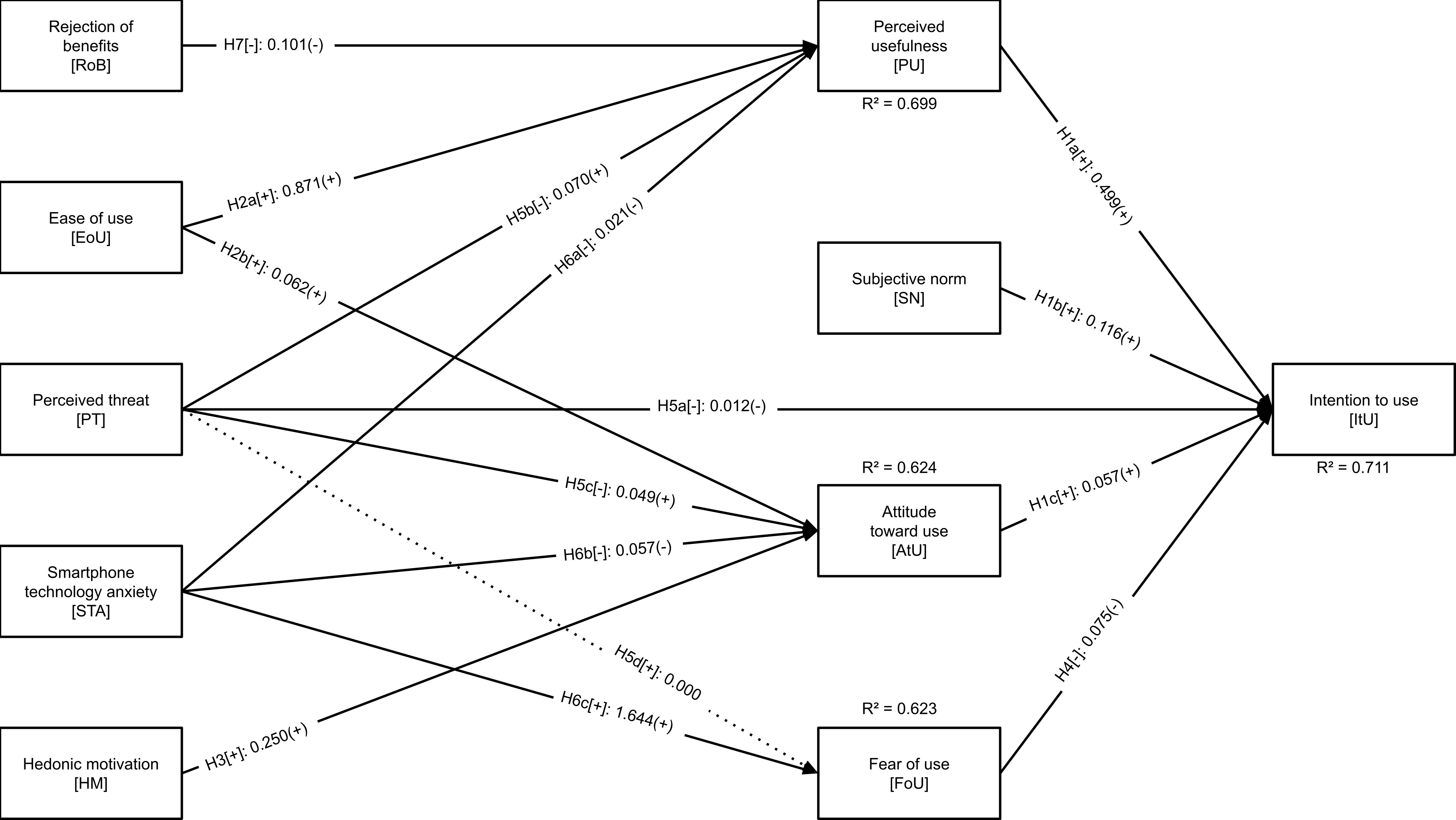}
  \caption{Full sample structural model. The values on associations are effect sizes $f^2$ ($small > 0.02, medium > 0.15, large > 0.35$). Dotted associations are non-significant ($p \geq 0.05$) while others are significant ($p < 0.05$).}
  \label{fig:sm}
\end{figure*}

We first applied the structural model to the full sample. Model fit indices indicate that the model fits the data well ($\chi^2 = 1,196.320$, $d. f. = 434$, $\chi^2 / d. f. = 2.756$, $CFI = 0.954$, $SRMR = 0.063$, $RMSEA = 0.057$). The results of hypothesis testing on the full sample are presented in Fig.~\ref{fig:sm}.

The structural model appears to explain very well all dependent variables with $R^2$ ranging from 0.623 to 0.711. The results indicate support for hypotheses H1a ($std. est. = 0.506$, $p < 0.001$), H1b ($std. est. = 0.217$, $p < 0.001$), H1c ($std. est. = 0.173$, $p < 0.001$), H2a ($std. est. = 0.643$, $p < 0.001$), H2b ($std. est. = 0.227$, $p < 0.001$), H3 ($std. est. = 0.454$, $p < 0.001$), H4 ($std. est. = -0.174$, $p < 0.001$), H5a ($std. est. = -0.064$, $p = 0.026$), H6a ($std. est. = -0.103$, $p = 0.011$), H6b ($std. est. = -0.188$, $p < 0.001$), and H6c ($std. est. = 0.791$, $p < 0.001$). Contrary to our expectations, the results indicate support for rejecting hypotheses H5b ($std. est. = 0.153$, $p < 0.001$) and H5c ($std. est. = 0.142$, $p < 0.001$). There was no support for hypothesis H5d ($std. est. = 0.035$, $p = 0.749$).

The association between ItU and PU has a large effect size while the effect sizes of associations of ItU with SN, AtU and FoU are small. Even though the association between ItU and PT is significant, its effect size appears to be negligible which diminishes support for hypothesis H5a. The association between PU and EoU has a large effect size while the associations of PU with STA, RoB and PT have small effect sizes. The association with PT is positive contrary to the posed hypothesis. The association between AtU and HM has a medium effect size while the effect sizes of the associations of AtU with EoU, STA and PT have small effect sizes. Again, the association with PT is positive contrary to the respective hypothesis. The association between FoU and STA has a large effect size while the effect size of the non-significant association between PT and FoU is negligible.

Next, we applied the structural model to the user and non-user subsamples. Model fit indices suggest that the model fits well the data from both subsamples. The results of hypothesis testing on user and non-user subsamples are presented in Table~\ref{tab:sm_user}.

\begin{table*}
  \caption{Structural model applied on user and non-user subsamples. Results are reported with 95\% CI.}
  \label{tab:sm_user}
  \centering
  \begin{tabular}{lllrr}
    \toprule
 &  &  & Users of smartphones & Non-users of smartphones \\ \midrule
$n$ &  &  & $313$ & $222$ \\
\textit{Model fit} &  &  &  &  \\
$\chi^2$ &  &  & $920.260$ & $765.801$ \\
$d. f.$ &  &  & $434$ & $434$ \\
$\chi^2/d. f.$ &  &  & $2.120$ & $1.765$ \\
CFI &  &  & $0.927$ & $0.934$ \\
SRMR &  &  & $0.079$ & $0.067$ \\
RMSEA &  &  & $0.060$ & $0.059$ \\
$R^2$ &  &  &  &  \\
ItU &  &  & $0.367 [0.284, 0.450]$ & $0.521 [0.433, 0.609]$ \\
PU &  &  & $0.484 [0.406, 0.562]$ & $0.511 [0.422, 0.600]$ \\
AtU &  &  & $0.445 [0.365, 0.525]$ & $0.447 [0.352, 0.541]$ \\
FoU &  &  & $0.594 [0.526, 0.663]$ & $0.419 [0.322, 0.516]$ \\
\textit{Effect size} &  &  &  &  \\
ItU & $\leftarrow$ & PU & $^{***}0.061 [0.009, 0.119]$ & $^{***}0.546 [0.343, 0.821]$ \\
 & $\leftarrow$ & SN & $^{***}0.092 [0.028, 0.166]$ & $^{***}0.108 [0.027, 0.204]$ \\
 & $\leftarrow$ & AtU & $^{**}0.033 [-0.005, 0.075]$ & $^{*}0.037 [-0.010, 0.090]$ \\
 & $\leftarrow$ & FoU & $^{***}0.058 [0.007, 0.115]$ & $0.006 [-0.014, 0.027]$ \\
 & $\leftarrow$ & PT & $0.003 [-0.009, 0.016]$ & $^{*}0.027 [-0.014, 0.072]$ \\
PU & $\leftarrow$ & EoU & $^{***}0.269 [0.154, 0.410]$ & $^{***}0.788 [0.529, 1.153]$ \\
 & $\leftarrow$ & PT & $^{**}0.057 [0.006, 0.113]$ & $^{*}0.046 [-0.007, 0.106]$ \\
 & $\leftarrow$ & STA & $^{***}0.067 [0.012, 0.128]$ & $0.009 [-0.015, 0.033]$ \\
 & $\leftarrow$ & RoB & $^{***}0.172 [0.082, 0.279]$ & $0.022 [-0.015, 0.062]$ \\
AtU & $\leftarrow$ & EoU & $0.006 [-0.011, 0.024]$ & $^{***}0.106 [0.025, 0.201]$ \\
 & $\leftarrow$ & HM & $^{***}0.214 [0.113, 0.336]$ & $^{***}0.160 [0.060, 0.282]$ \\
 & $\leftarrow$ & PT & $^{**}0.039 [-0.003, 0.085]$ & $^{**}0.041 [-0.009, 0.097]$ \\
 & $\leftarrow$ & STA & $^{***}0.153 [0.069, 0.252]$ & $0.005 [-0.013, 0.025]$ \\
FoU & $\leftarrow$ & PT & $0.002 [-0.008, 0.014]$ & $0.005 [-0.013, 0.024]$ \\
 & $\leftarrow$ & STA & $^{***}1.421 [1.074, 1.908]$ & $^{***}0.652 [0.421, 0.971]$ \\
    \bottomrule
  \end{tabular}
  \begin{flushleft}
  \textit{Notes}: $^* p < 0.05, ^{**} p < 0.01, ^{***} p < 0.001$; CFI -- comparative fit index, SRMR -- standardized root mean square residual, RMSEA -- root mean square error of approximation; SN -- subjective norm, PT -- perceived threat, HM -- hedonic motivation, EoU -- ease of use, STA -- smartphone technology anxiety, PU -- perceived usefulness, ItU -- intention to use, FoU -- fear of use, RoB -- rejection of benefits, AtU -- attitude toward use.
  \end{flushleft}
\end{table*}

Both structural models appear to explain well all dependent variables with $R^2$ ranging from 0.367 to 0.594 for the subsample of smartphone users, and $R^2$ ranging from 0.419 to 0.521 for non-users of smartphones. Nevertheless, $R^2$ for ItU, PU and AtU are significantly lower in both subsamples than in the full sample ($R_{ItU}^2 = 0.711$, 95\% CI $[0.670, 0.752]$, $R_{PU}^2 = 0.699$, 95\% CI $[0.656, 0.741]$, $R_{AtU}^2 = 0.624$, 95\% CI $[0.574, 0.674]$). Also, $R^2$ for FoU is significantly lower in the non-user subsample than in the full sample ($R_{FoU}^2 = 0.623$, 95\% CI $[0.573, 0.673]$). Compared to each other, the model explains FoU significantly better in the user subsample ($R^2 = 0.594$) than in the non-user subsample ($R^2 = 0.419$).

There are six differences in the significance of associations between both subsamples. The associations between ItU and FoU, PU and STA, PU and RoB, and AtU and STA are significant in the user subsample but not in the non-user subsample. The differences in effect size are however only significant for the associations between PU and RoB, and AtU and STA. The former association has a medium effect size in the user subsample and a small effect size in the non-user subsample. The latter association has a medium effect size in the user subsample while the effect size in the non-user subsample is negligible. The associations between ItU and PT, and AtU and EoU are significant in the non-user subsample but not in the user subsample. The differences in effect size are only significant for the latter association. Its effect size is small in the non-user subsample while the effect size in the user subsample is negligible.

There are three more significant differences in effect sizes for associations that are significant in both subsamples. First, the effect size of the association between ItU and PU is small in the user subsample and large in the non-user subsample. Second, the effect size of the association between PU and EoU is medium in the user subsample and large in the non-user subsample. Third, the effect size of the association between FoU and STA is large in both subsamples albeit larger in the user subsample which can be likely attributed to a significantly higher $R^2$ in the user subsample.

Finally, we applied the structural model to the subsamples according to awareness of smartphones. Model fit indices suggest that the model fits well the data from both subsamples. The results of hypothesis testing on aware and not aware subsamples are presented in Table~\ref{tab:sm_aware}.

\begin{table*}
  \caption{Structural model applied on aware and not aware subsamples. Results are reported with 95\% CI.}
  \label{tab:sm_aware}
  \centering
  \begin{tabular}{lllrr}
    \toprule
 &  &  & Aware of smartphones & Not aware of smartphones \\ \midrule
$n$ &  &  & $366$ & $169$ \\
\textit{Model fit} &  &  &  &  \\
$\chi^2$ &  &  & $1016.063$ & $722.653$ \\
$d. f.$ &  &  & $434$ & $434$ \\
$\chi^2/d. f.$ &  &  & $2.341$ & $1.665$ \\
CFI &  &  & $0.943$ & $0.926$ \\
SRMR &  &  & $0.077$ & $0.076$ \\
RMSEA &  &  & $0.061$ & $0.063$ \\
$R^2$ &  &  &  &  \\
ItU &  &  & $0.622 [0.562, 0.681]$ & $0.498 [0.396, 0.600]$ \\
PU &  &  & $0.598 [0.535, 0.660]$ & $0.559 [0.463, 0.655]$ \\
AtU &  &  & $0.583 [0.519, 0.647]$ & $0.417 [0.308, 0.526]$ \\
FoU &  &  & $0.597 [0.533, 0.660]$ & $0.483 [0.377, 0.588]$ \\
\textit{Effect size} &  &  &  &  \\
ItU & $\leftarrow$ & PU & $^{***}0.249 [0.146, 0.372]$ & $^{***}0.469 [0.263, 0.755]$ \\
 & $\leftarrow$ & SN & $^{***}0.126 [0.055, 0.208]$ & $^{**}0.060 [-0.008, 0.139]$ \\
 & $\leftarrow$ & AtU & $^{***}0.073 [0.019, 0.132]$ & $^{*}0.043 [-0.015, 0.108]$ \\
 & $\leftarrow$ & FoU & $^{***}0.103 [0.039, 0.175]$ & $0.024 [-0.019, 0.070]$ \\
 & $\leftarrow$ & PT & $0.001 [-0.006, 0.009]$ & $^{*}0.035 [-0.017, 0.094]$ \\
PU & $\leftarrow$ & EoU & $^{***}0.533 [0.370, 0.740]$ & $^{***}0.877 [0.567, 1.342]$ \\
 & $\leftarrow$ & PT & $^{***}0.071 [0.018, 0.130]$ & $^{*}0.068 [-0.005, 0.152]$ \\
 & $\leftarrow$ & STA & $^{*}0.026 [-0.006, 0.059]$ & $0.001 [-0.007, 0.009]$ \\
 & $\leftarrow$ & RoB & $^{***}0.135 [0.061, 0.220]$ & $0.021 [-0.020, 0.065]$ \\
AtU & $\leftarrow$ & EoU & $^{**}0.038 [-0.001, 0.079]$ & $^{**}0.061 [-0.008, 0.140]$ \\
 & $\leftarrow$ & HM & $^{***}0.339 [0.216, 0.490]$ & $^{***}0.121 [0.024, 0.240]$ \\
 & $\leftarrow$ & PT & $^{***}0.060 [0.012, 0.113]$ & $0.029 [-0.019, 0.082]$ \\
 & $\leftarrow$ & STA & $^{**}0.047 [0.004, 0.094]$ & $^{**}0.059 [-0.009, 0.137]$ \\
FoU & $\leftarrow$ & PT & $0.005 [-0.009, 0.019]$ & $0.008 [-0.018, 0.036]$ \\
 & $\leftarrow$ & STA & $^{***}1.467 [1.134, 1.924]$ & $^{***}0.880 [0.565, 1.352]$ \\
    \bottomrule
  \end{tabular}
  \begin{flushleft}
  \textit{Notes}: $^* p < 0.05, ^{**} p < 0.01, ^{***} p < 0.001$; CFI -- comparative fit index, SRMR -- standardized root mean square residual, RMSEA -- root mean square error of approximation; SN -- subjective norm, PT -- perceived threat, HM -- hedonic motivation, EoU -- ease of use, STA -- smartphone technology anxiety, PU -- perceived usefulness, ItU -- intention to use, FoU -- fear of use, RoB -- rejection of benefits, AtU -- attitude toward use.
  \end{flushleft}
\end{table*}

Both structural models explain well all dependent variables. $R^2$ ranged from 0.583 to 0.622 in the aware subsample. These were not significantly lower than $R^2$ in the full sample. In the not aware subsample, $R^2$ ranged from 0.417 to 0.559. $R^2$ for ItU, PU and AtU are significantly lower than in the full sample ($R_{ItU}^2 = 0.711$, 95\% CI $[0.670, 0.752]$, $R_{PU}^2 = 0.699$, 95\% CI $[0.656, 0.741]$, $R_{AtU}^2 = 0.624$, 95\% CI $[0.574, 0.674]$). Nevertheless, there were no significant differences between the two subsamples when compared to each other.

There are five differences in the significance of associations between the two subsamples. The associations between ItU and FoU, PU and STA, PU and RoB, and AtU and PT are only significant in the aware subsample. The association between ItU and PT is however only significant in the not aware subsample. There were no significant differences in effect sizes though.
We present a summary of hypothesis testing results in Table~\ref{tab:summary}.

\begin{table*}
  \caption{Summary of hypothesis testing results}
  \label{tab:summary}
  \centering
  \begin{tabular}{lL{.68\textwidth}L{.09\textwidth}}
    \toprule
Hypothesis & Evidence & Result \\ \midrule
H1a[+] PU $\rightarrow$ ItU & • Significant positive association, large effect size (full sample, non-user and not aware subsamples) & Supported \\
 & • Significant positive association, small effect size (user subsample) &  \\
 & • Significant positive association, medium effect size (aware subsample) &  \\
H1b[+] SN $\rightarrow$ ItU & • Significant positive association, small effect size & Supported \\
H1c[+] AtU $\rightarrow$ ItU & • Significant positive association, small effect size & Supported \\
H2a[+] EoU $\rightarrow$ PU & • Significant positive association, large effect size (full sample, non-user, aware and not aware subsamples) & Supported \\
 & • Significant positive association, medium effect size (user subsample) &  \\
H2b[+] EoU $\rightarrow$ AtU & • Significant positive association, small effect size (full sample, non-user, aware and not aware subsamples) & Partially supported \\
 & • Non-significant association (user subsample) &  \\
H3[+] HM $\rightarrow$ AtU & • Significant positive association, medium effect size (full sample, user, non-user and aware subsamples) & Supported \\
 & • Significant positive association, small effect size (not aware subsample) &  \\
H4[-] FoU $\rightarrow$ ItU & • Significant negative association, small effect size (full sample, user and aware subsamples) & Partially supported \\
 & • Non-significant association (non-user and not aware subsamples) &  \\
H5a[-] PT $\rightarrow$ ItU & • Significant negative association, negligible effect size (full sample) & Partially supported \\
 & • Significant negative association, small effect size (non-user and not aware subsamples) &  \\
 & • Non-significant association (user and aware subsamples) &  \\
H5b[-] PT $\rightarrow$ PU & • Significant positive association, small effect size & Rejected \\
H5c[-] PT $\rightarrow$ AtU & • Significant positive association, small effect size (full sample, user, non-user and aware subsamples) & Partially rejected \\
 & • Non-significant association (not aware subsample) &  \\
H5d[+] PT $\rightarrow$ FoU & • Non-significant association & Not supported \\
H6a[-] STA $\rightarrow$ PU & • Significant negative association, small effect size (full sample, user and aware subsamples) & Partially supported \\
 & • Non-significant association (non-user and not aware subsamples) &  \\
H6b[-] STA $\rightarrow$ AtU & • Significant negative association, small effect size (full sample, aware and not aware subsamples) & Partially supported \\
 & • Significant negative association, medium effect size (user subsample) &  \\
 & • Non-significant association (non-user subsample) &  \\
H6c[+] STA $\rightarrow$ FoU & • Significant positive association, large effect size & Supported \\
H7[-] RoB $\rightarrow$ PU & • Significant negative association, small effect size (full sample and aware subsample) & Partially supported \\
 & • Significant negative association, medium effect size (user subsample) &  \\
 & • Non-significant association (non-user and not aware subsamples) &  \\
    \bottomrule
  \end{tabular}
  \begin{flushleft}
  \textit{Notes}: SN -- subjective norm, PT -- perceived threat, HM -- hedonic motivation, EoU -- ease of use, STA -- smartphone technology anxiety, PU -- perceived usefulness, ItU -- intention to use, FoU -- fear of use, RoB -- rejection of benefits, AtU -- attitude toward use.
  \end{flushleft}
\end{table*}

\section{Discussion}
\label{sec:discussion}

\subsection{Theoretical and practical implications}

This study makes several theoretical contributions. First, this study provides new insights into adoption of smartphones by older adults. Even though adoption of smartphones by older adults has been studied extensively in existing literature, this study incorporates new associations into the theoretical framework. The results of this study confirmed associations of ItU with PU and SN further supporting the findings found in the existing literature \cite{conci_useful_2009,huang_using_2023,yang_acceptance_2023}. The results also provide additional support for literature that indicates the association between ItU and AtU should be included in addition to the association between ItU and PU in technology adoption studies among older adults \cite{nayak_application_2010,yeh_effect_2024}. This is in line with the original TAM \cite{davis_user_1989}, and contrary to its later versions, such as the revised TAM \cite{davis_user_1989} and TAM2 \cite{venkatesh_theoretical_2000}. The results of this study also confirm the existing literature by providing support for associations of PU with EuO \cite{conci_useful_2009,huang_using_2023}, and STA \cite{huang_using_2023,zhu_staying_2024}. The results of this study further support the current literature by indicating support for associations of AtU with EoU \cite{chen_gerontechnology_2014,guner_use_2020,jarvis_technology_2020}, and STA \cite{ellis_why_2021}. AtU is also associated with HM supporting the literature indicating that playfulness (a conceptually related construct) is associated with AtU \cite{huang_what_2023}.

For the purposes of this study, we uncovered an association between RoB and PU albeit with a small effect size. This association is also non-significant in the non-user and not aware subsamples indicating that rejection of benefits plays a more important role for users of smartphones with a medium effect size, and older adults who are aware of smartphones even though with a small effect size. The results of this study also indicate that FoU plays an important role in adoption of smartphones by older adults albeit with a small effect size. Similarly to the association between RoB and PU, this association is also non-significant in the non-user and not aware subsamples indicating that fear of using smartphones plays a more important role for users of smartphones and older adults who are aware of smartphones. We finally found that FoU is associated with STA suggesting that older adults fear the use of smartphones due to their technological anxiety. These findings contribute to the literature on adoption of smartphones among older adults. Future studies on adoption of smartphones among older adults may thus include RoB and FoU in their research models.

Second, this study is one of the first to study the role of perceived threat of cyberattacks in adoption of smartphones by older adults. The significant negative association between ItU and PT in the full sample, and the non-user and not aware subsamples provides some support for the existing literature which predicts either a direct \cite{lai_design_2016} or indirect \cite{conci_useful_2009,lai_design_2016} association between the two. This is also in line with the protection motivation theory (PMT) which posits that higher perceived threat (e.g., of cyberattacks) results in higher protection motivation (e.g., avoidance of smartphone use) \cite{burns_examining_2017,vrhovec_explaining_2023}. The negligible effect size of this association indicates that perceived threat of cyberattacks in the full sample and non-significant association in the user and aware subsamples suggest that PT plays a more important role in adoption of smartphones for non-users of smartphones and older adults who are less aware of them.

PT appears to be associated with both PU and AtU. However, surprisingly these associations are positive which is contrary to what we hypothesized. We assumed that older adults who perceive a higher threat of cyberattacks would perceive less usefulness and would have a poorer attitude toward use. The results however indicate that this is not the case. This is also not in line with existing literature which predicts a negative association between PU and PT \cite{conci_useful_2009,lai_design_2016}. A possible explanation for positive associations could be awareness. Older adults with a better attitude toward the use of smartphones and who are more aware of their usefulness may be also more aware of cyberthreats. We could also switch the perspective by saying that older adults who feel less threatened by cyberattacks are also less likely to grasp the usefulness of smartphones and have poorer attitude towards their use due to their poor awareness of the abilities of smartphones and cyberthreats. In our study, we measured awareness of smartphones which might be related to other kinds of awareness mentioned before. Even though we found no significant differences between respondents who were aware of smartphones and those who are not, the association between AtU and PT is non-significant in the not aware subsample indicating some support for this potential explanation. Finally, it is surprising that the association between FoU and PT was non-significant suggesting that older adults do not fear using smartphones because they would feel threatened by cyberattacks. This finding is consistent with the assumption that the unexpected positive associations of PT with PU and AtU may be due to poor awareness of cyberthreats. Overall, the role of PT, and more broadly cybersecurity, in adoption of smartphones by older adults has not been sufficiently explained yet. Future studies, such as studies including other kinds of awareness, would be needed to provide further insights.

Third, this study is one of the first to study differences between users and non-users of smartphones. The results of our study indicate that PU plays a more important role in adoption of smartphones by non-users of smartphones. EoU plays similarly a more important role in shaping PU and AtU among non-users. Even though these associations have been confirmed before, this study is among the first to suggest the greater importance of EoU among non-users of smartphones therefore contributing to the literature on adoption of smartphones by older adults. Contrary, STA appears to play a more important role in shaping PU and AtU among users of smartphones. This study is among the first to indicate the greater importance of STA among users of smartphones which may seem counter-intuitive. Studies however suggest that STA is associated with EoU \cite{cimperman_analyzing_2016} thus indirectly affecting adoption of smartphones. This explanation is consistent with our findings. The studied constructs explain significantly better FoU in the user than in the non-user subsample. Since STA is the only significant construct associated with FoU, these results suggest the existence of other related factors that were not accounted for in our research model. Future studies may explore additional factors that shape the fear of using smartphones among older adults who are not using smartphones.

Finally, this study has some practical implications. It indicates that campaigns promoting adoption of smartphones among older adults should primarily emphasize the usefulness of smartphones since it has the largest effect size among all constructs associated with ItU. Among non-users of smartphones, EoU plays the strongest role in shaping both PU and AtU and could thus be emphasized to achieve this outcome. Among smartphone users, efforts to lower STA may achieve the same. As older adults age, their abilities generally decrease, and they may feel more anxiety as their age progresses. Regular workshops, training and informal teaching and/or refreshing of knowledge and skills may help to keep their STA low or lower it.

A practical implication may also stem from the surprising rejection of hypotheses H5b and H5c indicating positive associations of PT with PU and AtU, respectively. We assumed that raising the awareness of cyberthreats could hinder the adoption of smartphones among older adults. These results however suggest the opposite. Incorporating cybersecurity materials and/or training in campaigns promoting adoption of smartphones among older adults, workshops, training and informal teaching may have a side-effect of indirectly improving adoption of smartphones. The readers should nevertheless exercise caution when implementing such improvements since more research is needed to further confirm our findings, possibly involving these activities.

\subsection{Limitations and future work}

This study has some limitations that the readers should note. First, the sample is not representative since the respondents in our sample seem to have higher formal education than the population of older adults in Slovenia. Notably, older adults with the lowest levels of formal education are underrepresented in the sample. Caution should be exercised when generalizing the results of our study to the population of older adults in Slovenia. Next, the study was conducted in a single cultural context. Even though the results may be applicable to similar cultural contexts, such as other EU countries, they may not be universally generalizable. Future studies comparing the results in varying cultural contexts may thus be beneficial in determining whether the findings are applicable to other cultural contexts. Finally, this study provides some insights into the role of awareness of smartphones in explaining the surprising positive associations of PT with PU and AtU. Future studies measuring other kinds of awareness, such as awareness of cyberthreats, or other factors may help to better explain these.

\begin{acks}
\small
This work was partially funded by the Slovenian Research and Innovation Agency [grant number J5-3111] and [grant number L5-50163]. The funders had no role in study design, data collection and analysis, decision to publish, or preparation of the manuscript.
\end{acks}

\bibliographystyle{ACM-Reference-Format}


\end{document}